\documentclass[aps,prl,amsmath,amssymb,floatfix,twocolumn,amsmath,superscriptaddress,twocolumn,nofootinbib,tighten,letterpaper]{revtex4}
\usepackage{multirow}
\usepackage{bbold}
\usepackage{subfigure}
\usepackage{color}
\usepackage{mathrsfs}
\usepackage{hyperref}
\usepackage[normalem]{ulem}
\usepackage{bm}

\usepackage{amsfonts, relsize, color}
\usepackage{graphics}
\usepackage{graphicx}
\usepackage{subfigure}
\usepackage{hyperref}
\usepackage{color}
\usepackage{comment}

\renewcommand\vec[1]{\ensuremath\boldsymbol{#1}} 

\begin{document}
\title{\bf Projected Topological Branes}

\author{Archisman Panigrahi}
\affiliation{Indian Institute of Science, Bangalore 560012, India}
\affiliation{Max-Planck-Institut f\"{u}r Physik komplexer Systeme, N\"{o}thnitzer Str. 38, 01187 Dresden, Germany}

\author{Vladimir Juri\v ci\' c}\thanks{Corresponding author:vladimir.juricic@su.se}
\affiliation{Nordita, KTH Royal Institute of Technology and Stockholm University,
Hannes Alfv\' ens v\" ag 12, SE-106 91 Stockholm, Sweden}
\affiliation{Departamento de F\'isica, Universidad T\'ecnica Federico Santa Mar\'ia, Casilla 110, Valpara\'iso, Chile}

\author{Bitan Roy}\thanks{Corresponding author:bitan.roy@lehigh.edu}
\affiliation{Max-Planck-Institut f\"{u}r Physik komplexer Systeme, N\"{o}thnitzer Str. 38, 01187 Dresden, Germany}
\affiliation{Department of Physics, Lehigh University, Bethlehem, Pennsylvania, 18015, USA}

\date{\today}

\maketitle

\noindent
{\bf Abstract}\\
{\bf Nature harbors crystals of dimensionality ($d$) only up to three. Here we introduce the notion of  \emph{projected topological branes} (PTBs): Lower-dimensional branes embedded in higher-dimensional parent topological crystals, constructed via a geometric cut-and-project procedure on the Hilbert space of the parent lattice Hamiltonian. When such a brane is inclined at a rational or an irrational slope, either a new lattice periodicity or a quasicrystal emerges. The latter gives birth to topoquasicrystals within the landscape of PTBs. As such PTBs are shown to inherit the hallmarks, such as the bulk-boundary, bulk-dislocation correspondences and topological invariant, of the parent topological crystals. We exemplify these outcomes by focusing on two-dimensional parent Chern insulators, leaving its signatures on projected one-dimensional (1D) topological branes in terms of localized endpoint, dislocation modes and the local Chern number. Finally, by stacking 1D projected Chern insulators, we showcase the imprints  of three-dimensional Weyl semimetals in $d=2$, namely the Fermi arc surface states and bulk chiral zeroth Landau level, responsible for the chiral anomaly. Altogether, the proposed PTBs open a realistic avenue to harness higher-dimensional ($d>3$) topological phases in laboratory.
}

\noindent
{\bf Introduction}\\
Crystals foster periodic structures, characterized by a set of discrete symmetries, such as translations, rotations and reflections, with their dimensionality limited to three in nature. In crystalline materials electrons experience periodic potential generated by the underlying periodic lattice, which in turn yields a rich manifold of electronic band structures. The latter can exhibit nontrivial topological properties, giving birth to topological crystals with remarkably universal features, such as boundary gapless modes, encoding the hallmark bulk-boundary correspondence~\cite{TKNN1982,haldane-prl1988,kane-mele-prl2005,bhz-science2006,molenkamp-science2007,hsieh-nature2008,xia-natphys2009,hsieh-science2010, chen-science2010,xu-science2015,lv-prx2015,hasan-kane-rmp2010,qi-zhang-rmp2011}. By now the role of crystalline symmetries has been explored exhaustively, giving rise to general classifications of insulating and gapless topological states, as well as their material realizations, which are naturally limited to three spatial dimensions~\cite{fu-prl2011,dziawa-natmat2012,slager-natphys2013,sato-PRB2014,kruthoff-prx2017,bradlyn-nature2017,zhang-nature2019, vergniory-nature2019,tang-nature2019}. However, the pursuit of extending the jurisdiction of  topological crystals beyond three spatial dimensions in real materials thus far remained elusive.

\begin{figure}[t!]
\includegraphics[width=0.95\linewidth]{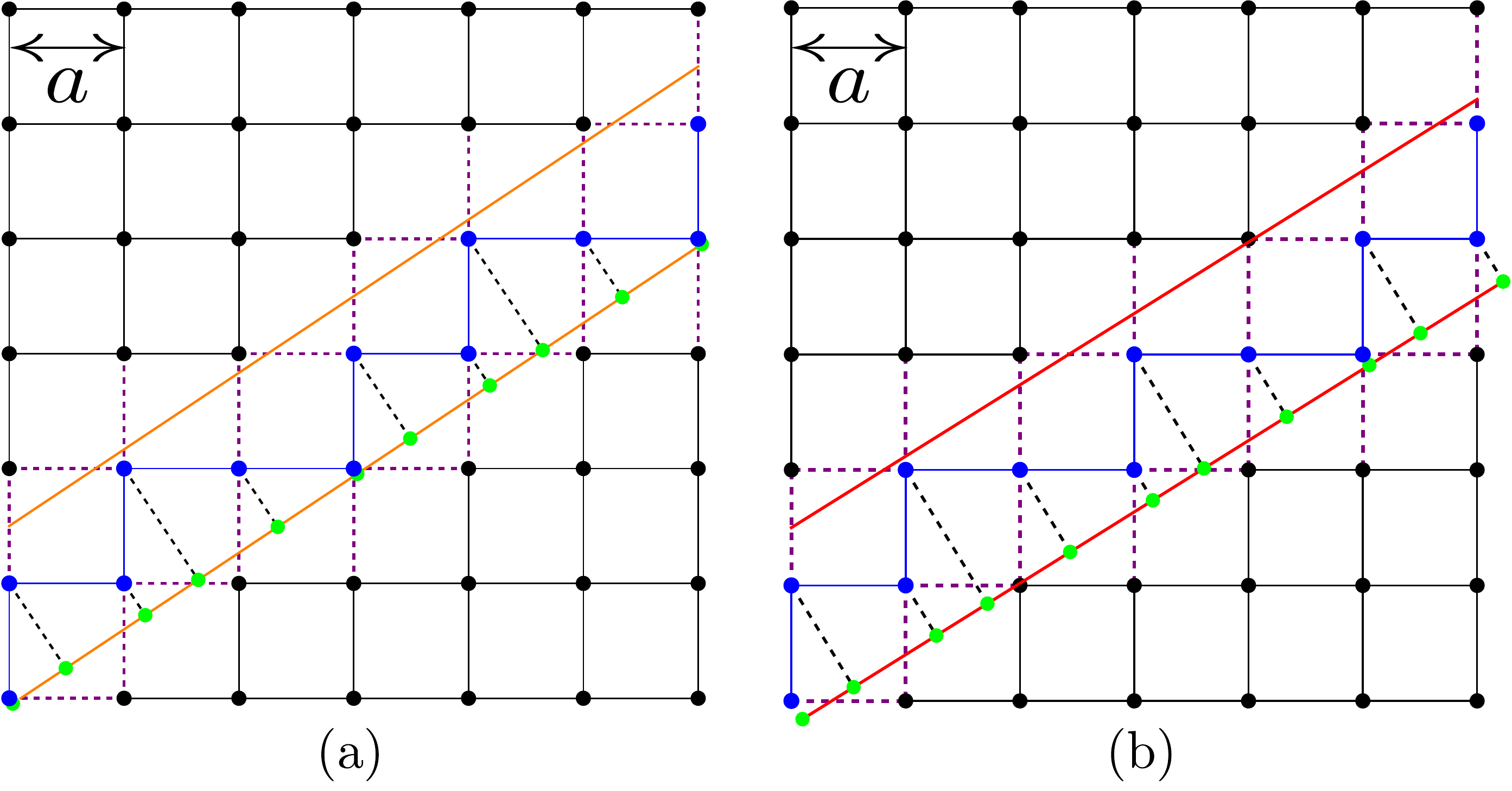}
\caption{{\bf Construction of a one-dimensional (1D) projected lattice}. The projected lattice is constituted by the sites residing within the (a) orange and (b) red lines, from a parent two-dimensional (2D) square lattice. These lines are defined by $y_i=S (x-\tilde{x}_i)+x_i$ for $i=u$ and $d$, with $x_{u}>x_{d}$, $\tilde{x}_u=1$ and $\tilde{x}_d=2$. Here $x$ measures the horizontal coordinate of lattice sites. In (a) $S=2/3$ and in (b) $S=\varphi^{-1}$, where $\varphi=(1+\sqrt{5})/2$ is the golden ratio. Notice $2/3$ is a rational approximant of $\varphi^{-1}$. When the garnered sites (blue dots) are projected (shown by black dashed lines) onto a 1D chain (green dots), a new lattice periodicity emerges in (a), while in (b) they constitute a Fibonacci quasicrystal, devoid of periodicity. Thermodynamic limit can be approached by increasing $x_u$ for a fixed $x_d$, which does not alter the periodicity in (a) or Fibonacci sequence in (b), such that the number of sites within the brane $N/L^2 \to 0$, where $L$ is the linear dimension of the parent square lattice in $x$ and $y$ directions, as both $N, L \to \infty$. The solid black (blue) lines constitute the hopping matrix elements of $H_{22}$ ($H_{11}$) and purple dashed lines to $H_{12}$ and $H_{21}$ appearing in $H_{\rm PTB}$ [Eq.~(\ref{eq:generalHamil})], originating from a parent Hamiltonian containing only nearest-neighbor hopping elements [Eq.~(\ref{eq:hamilchern})].
}~\label{fig:projectiongeometry}
\end{figure}

\begin{figure*}[t!]
\includegraphics[width=0.95\linewidth]{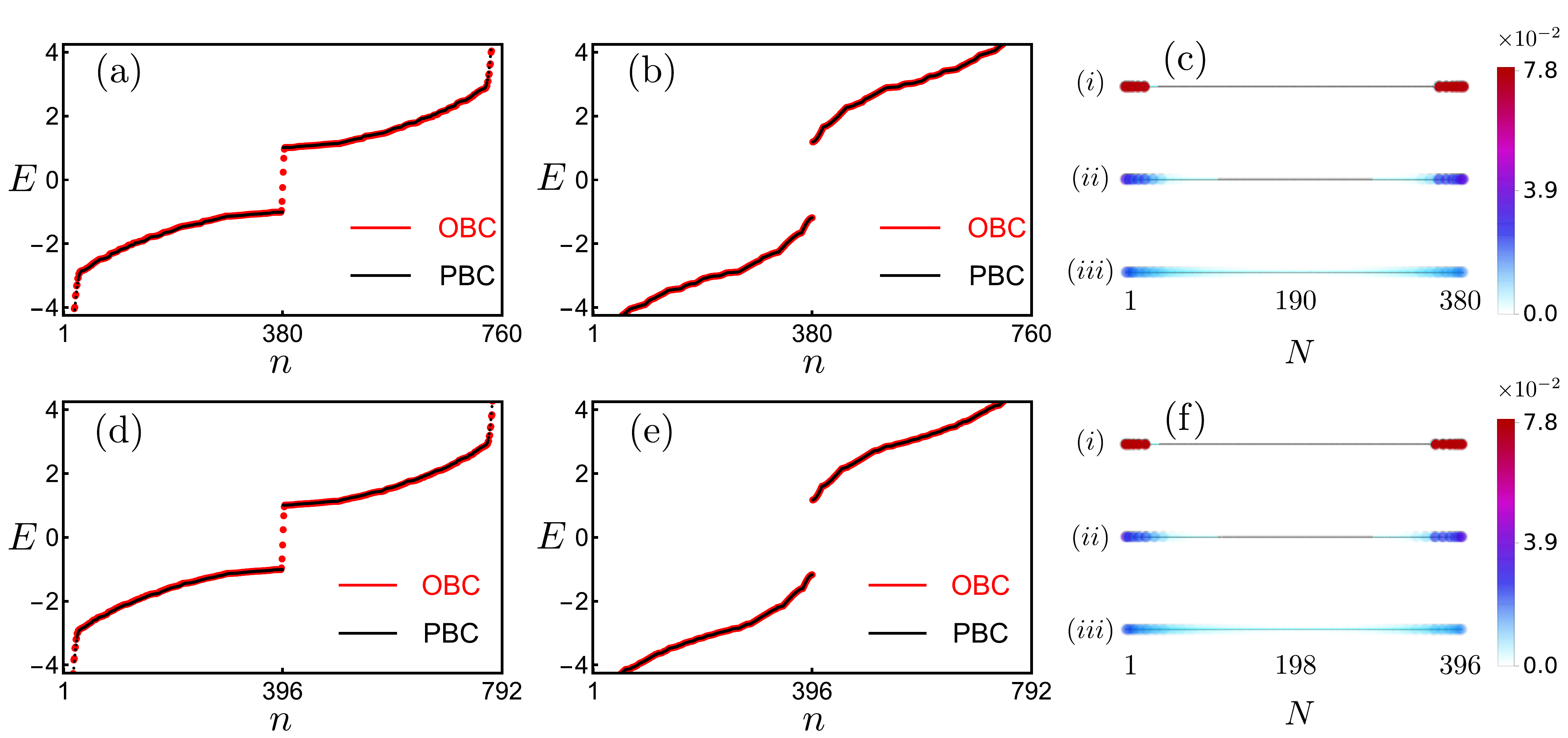}
\caption{{\bf Bulk-boundary correspondence of a projected Chern insulator in one dimension}. The underlying 1D chain is embedded in a parent 2D square lattice at a rational [(a)-(c)] and an irrational [(d)-(f)] slopes. Energy spectra of $H_{\rm PTB}$ [Eq.~(\ref{eq:generalHamil})] for (a) $m/t_0=2$ and $6$ (yielding identical energy spectra), and (b) $m/t_0=-2$ and $10$ (yielding identical energy spectra) with periodic (black) and open (red) boundary conditions. Here we set $t=2 t_0=1$ [Eq.~(\ref{eq:hamilchern})]. Only in the topological regime [panel (a)] we find near zero-energy mid-gap modes in systems with open boundary condition. (c) The local density of states of these modes are highly localized at the endpoints of the 1D chain hosting projected Chern insulators, as shown in $(i)$ for $m/t_0=2$ and $6$. As we approach the band gap closing, these modes start to delocalize as shown in $(ii)$ for $m/t_0=3.5$ and $4.5$, and $(iii)$ for $m/t_0=3.8$ and $4.2$. Panels (d)-(f) are analogous to panels (a)-(c), respectively. Here we consider a parent square lattice with linear dimension $L=60$ in both $x$ and $y$ directions, and 1D chains are constructed with $x_{\rm u}=12$ and $x_{\rm d}=6$ [Fig.~\ref{fig:projectiongeometry}], such that it contains only $10.5\%$ [(a)-(c)] and $11.0\%$ [(d)-(f)] sites of the parent crystal.
}~\label{fig:endpointLDOS}
\end{figure*}

\begin{figure}[t!]
\includegraphics[width=0.95\linewidth]{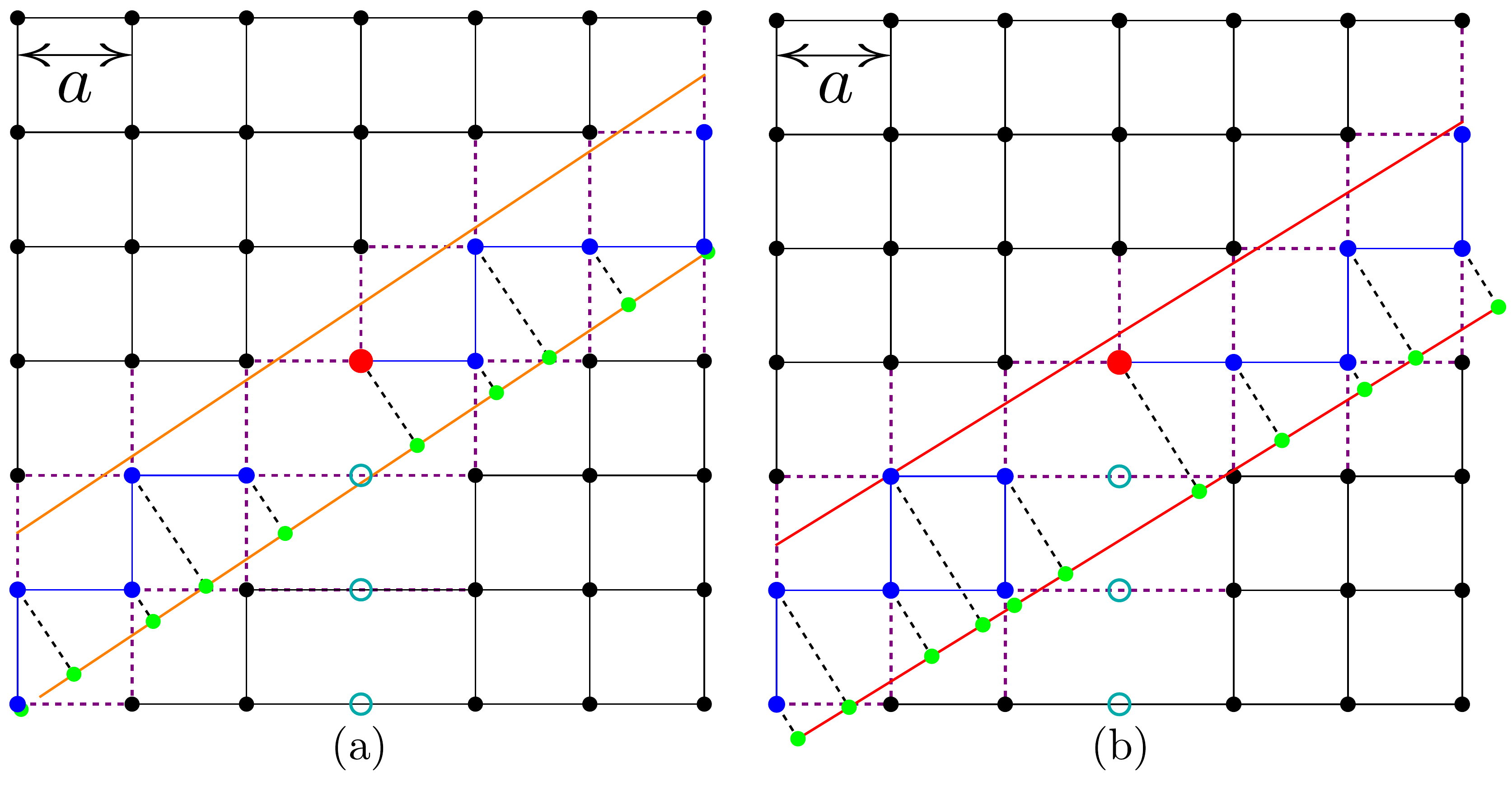}
\caption{{\bf Construction of a single edge dislocation}. Volterra cut-and-paste procedure on a parent 2D square lattice by removing a line of atoms (cyan circles), ending at its center or core (red circle) and subsequently reconnecting the sites living on the edges across it. Projected (a) 1D crystal and (b) 1D quasicrystal in the presence of a dislocation, with its core falling within the brane. The Burgers vector ${\bf b}= a {\bf e}_x$ in both cases resides within the hyperplane of the parent crystal. Rest of the details are the same as in Fig.~\ref{fig:projectiongeometry}.
}~\label{fig:dislocationgeometry}
\end{figure}

\begin{figure*}[t!]
\includegraphics[width=0.95\linewidth]{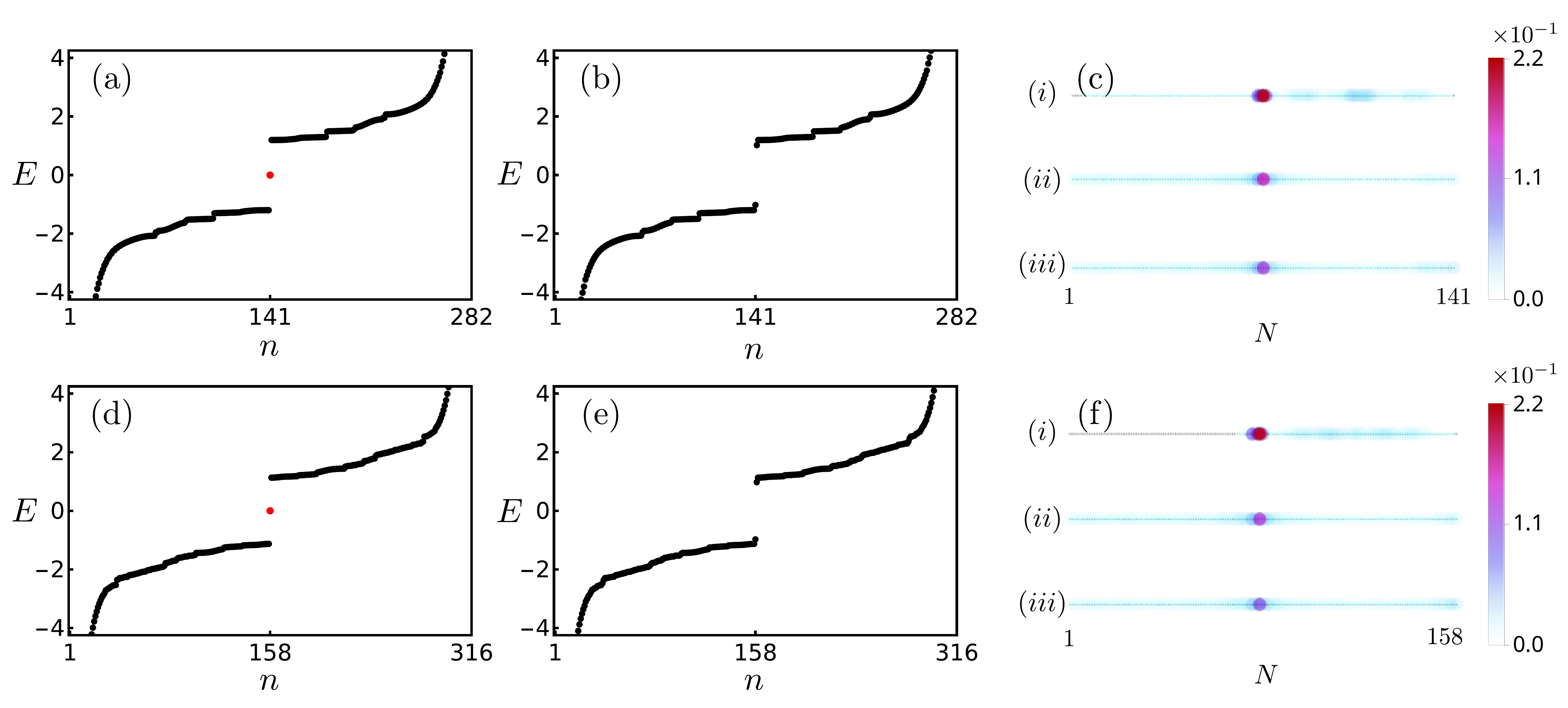}
\caption{{\bf Single edge dislocation with Burgers vector ${\bf b}= a {\bf e}_x$ [Fig.~\ref{fig:dislocationgeometry}] as a bulk probe of translationally active projected Chern insulator in one dimension}. The underlying 1D chain is embedded in a 2D square lattice with rational [(a)-(c)] and irrational [(d)-(f)] slopes. Zero energy dislocation modes (in red) can be found in the spectra of $H_{\rm PTB}$ [Eq.~(\ref{eq:generalHamil})] only for (a) the ${\rm M}$ phase ($m/t_0=6$), but not for (b) the $\Gamma$ phase ($m/t_0=2$) of the parent 2D Chern insulator. Here we take $t=2 t_0=1$ [Eq.~(\ref{eq:hamilchern})], and impose periodic boundary condition in the $x$ direction. The linear dimension of the parent square lattice is $L=60$ in both $x$ and $y$ directions, and dislocation core at $(31 a, 30 a)$ falls within the projected 1D chain. Such a chain is constructed with $x_u=12$ and $x_d=10$ [(a)-(c)], and $x_u=14$ and $x_d=12$ [(d)-(f)], respectively containing $7.8\%$ and $8.7\%$ of the sites from parent square lattice [Fig.~\ref{fig:projectiongeometry}]. (c) The local density of states for the zero-energy dislocation modes are strongly localized at the core of the dislocation, after being projected onto the 1D chain, for $(i)$ $m/t_0=6$. As we approach the band gap closing these modes delocalize as shown in $(ii)$ for $m/t_0=4.02$ and $(iii)$ for $m/t_0=4.01$, which can be seen from the prominent gradual decay of the local density of states for the dislocation mode at its core. In addition, the local density of states spreads significantly over the entire 1D brane as we approach the band gap closing point at $m/t_0=4$, also indicating the melting of the dislocation mode. A tiny fraction of the spectral weight for zero energy mode appears on the right side of the 1D brane even for $m/t_0=6.0$ due to the leakage of dislocation mode to the $y$-directional edges in the parent square lattice, as we impose the open boundary condition in the $y$ direction with a single edge dislocation with the Burgers vector ${\bf b}= a {\bf e}_x$. A weak left-right asymmetry arises since the 1D brane breaks such symmetry in the parent square lattice [Fig.~\ref{fig:dislocationgeometry}]. Panels (d)-(f) are analogous to panels (a)-(c), respectively.
}~\label{fig:dislocationLDOS}
\end{figure*}

\begin{figure}[t!]
\includegraphics[width=0.95\linewidth]{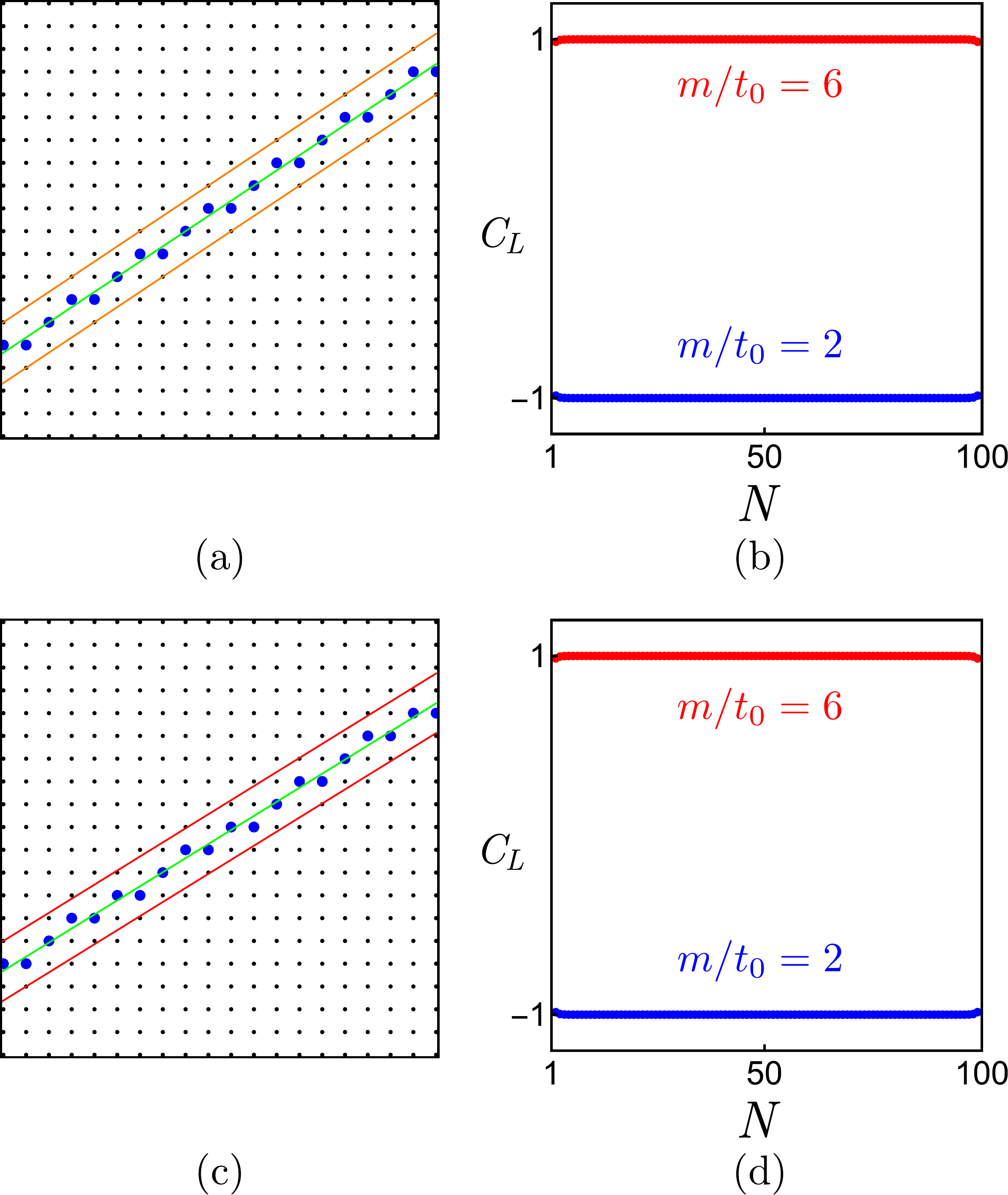}
\caption{{\bf Topological invariant of projected Chern insulators on 1D branes}. (a) Local Chern number ($C_L$) is computed on the blue sites, residing around the green lines, given by $y_{\rm mid}=(y_u+y_d)/2$ when the brane is inclined at a rational slope [Fig.~\ref{fig:projectiongeometry}]. (b) The variation of $C_L$ along the green line, on which the sites are indexed by $N$, is shown for $m/t_0=2$ (blue dots) and $m/t_0=6$ (red dots) [Eq.~(\ref{eq:hamilchern})], displaying its quantized values, equal to the Chern number (C) of parent 2D Chern insulator. Panels (c) and (d) are analogous to panels (a) and (b), respectively, but when the brane is inclined at an irrational slope. For details see Supplementary Note 1 and Supplementary Figure 1 of the SI. The results are obtained by constructing the branes from a parent square lattice of linear dimension $L=100$ in both $x$ and $y$ direction for $x_u=22$ and $x_d=12$, such that they contain (b) $10.3 \%$ and (d) $10.6 \%$ of sites of parent square lattice [Fig.~\ref{fig:projectiongeometry}].
}~\label{fig:topoinvariant}
\end{figure}

\begin{figure*}[t!]
\includegraphics[width=0.95\linewidth]{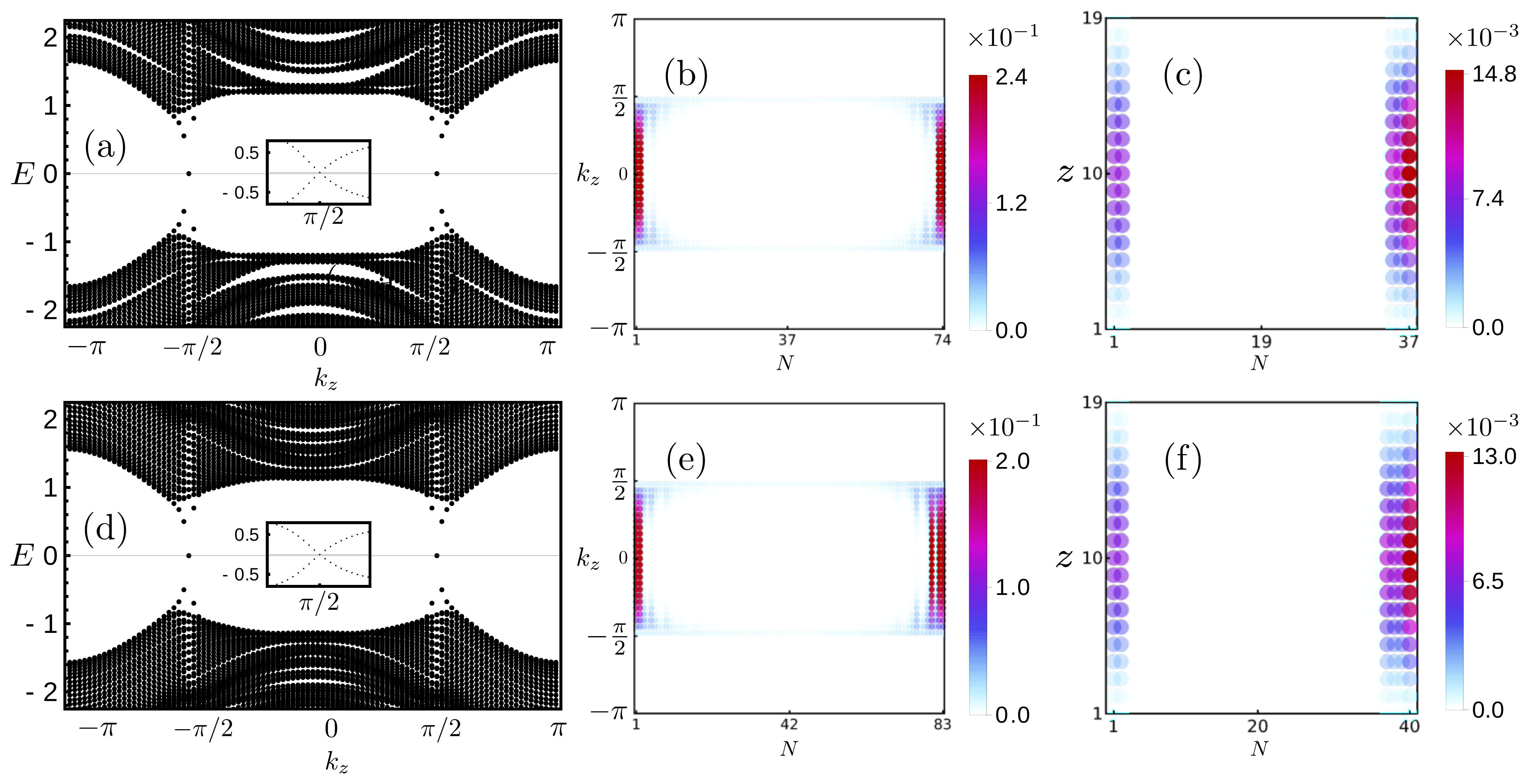}
\caption{{\bf Projected Weyl semimetal (PWSM)}. PWSM is constructed by stacking 1D projected Chern insulator in the $z$ direction in a translationally invariant fashion. The underlying 1D chain is inclined at a rational [irrational] slope in the parent 2D square lattice in panels (a)-(c) [panels (d)-(f)] row. (a) Energy spectra of $H_{\rm PTB}(k_z)$ [Eq.~(\ref{eq:generalHamil})] for $m=4$ and $t=t_z=2t_0=1$ [Eqs.~(\ref{eq:hamilchern}) and (\ref{eq:weakcoupling})], displaying two Weyl nodes at $k_z=\pm \pi/2$ (see Inset). Here $k_z$ is measured in units of $a^{-1}$ and we impose periodic boundary condition in both $x$ and $y$ direction. (b) Fermi arc states of $H_{\rm PTB}(k_z)$ in between two Weyl nodes with open boundary conditions in the $x$ and $y$ directions. As we approach the Weyl nodes, the Fermi arcs get delocalized, and at $k_z=\pm \pi/2$ they leak through the bulk to connect two ends of the underlying projected 1D chain. For (a) and (b) the linear dimension of the square lattice is $L=32$ in both $x$ and $y$ directions, and the 1D chain is constructed with $x_u=8$ and $x_d=6$, such that it contains $19.7\%$ [(a)-(c)] and $20.0\%$ [(d)-(f)] of the sites of parent square lattice [Fig.~\ref{fig:projectiongeometry}]. (c) Fermi arcs with open boundary conditions along all three directions in a system with linear dimensions $L=16$ in the $x$ and $y$ directions, and $L=19$ in the $z$ direction. The 1D brain for each $z$ is constructed with $x_u=7$ and $x_d=5$ in panels (a)-(c), and $x_u=8$ and $x_d=6$ in panels (d)-(f), respectively containing $14.5\%$ and $15.6\%$ of the sites of the parent 3D cubic lattice. The local density of states shows left-right asymmetry as the embedded 1D chain breaks such symmetry in the parent 2D square lattice [Fig.~\ref{fig:projectiongeometry}]. Panels (d)-(f) are analogous to panels (a)-(c), respectively.
}~\label{fig:projectedWSM}
\end{figure*}

\begin{figure}[t!]
\includegraphics[width=0.95\linewidth]{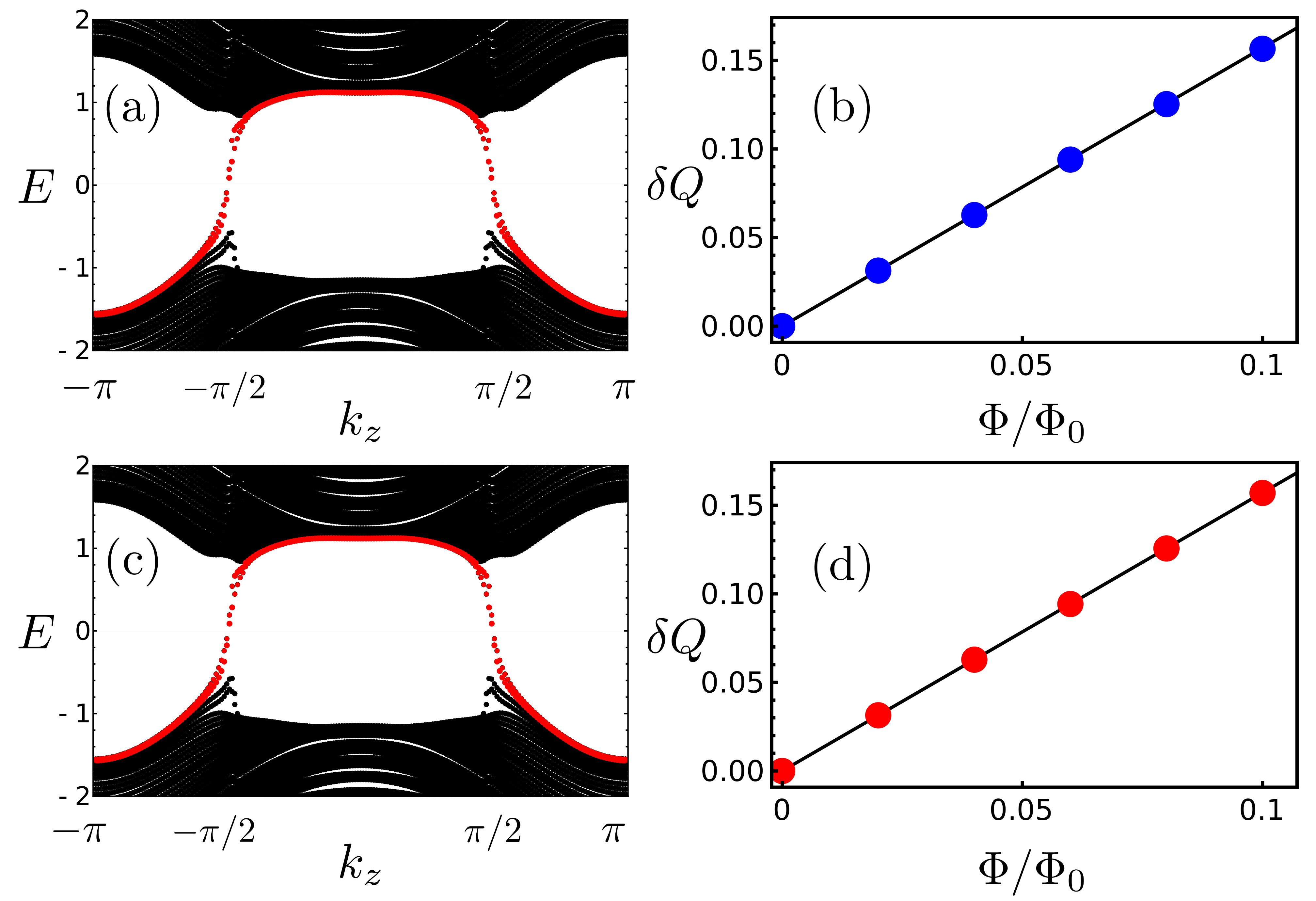}
\caption{{\bf Chiral zeroth Landau level and chiral anomaly in a projected Weyl semimetal (PWSM)}. The underlying 1D chain (hosting projected Chern or normal insulators) is embedded in a parent 2D square lattice at a rational [(a) and (b)] and an irrational [(c) and (d)] slopes. (a) Energy spectra of $H_{\rm PTB}(k_z, {\bf B})$ [Eq.~(\ref{eq:generalHamil})] in the presence of a uniform magnetic field ${\bf B}=B \hat{z}$ with periodic (open) boundary condition along $x \; (y)$ direction. Chiral zeroth Landau level (red dots) crosses the zero energy at $k_z=\pm \pi/2$, locations of the Weyl nodes [Fig.~\ref{fig:projectedWSM}]. (b) Accumulated charge ($\delta Q$) around a magnetic flux tube piercing a PWSM in the $z$ direction as a function of $\Phi/\Phi_0$, where $\Phi$ ($\Phi_0$) is the magnetic flux (flux quantum). Within the numerical accuracy the slope of the straight line is $\pi/2$, in agreement with field theoretic prediction~\cite{vazifeh-franz, dantas-banitez-roy-surowka}. Panels (c) and (d) are analogous to panels (a) and (b), respectively. The linear dimension of the system $L=32$ in the $x$ and $y$ directions, and we set $x_u=8$ and $x_d=6$ [Fig.~\ref{fig:projectiongeometry}].
}~\label{fig:Chiralanomaly}
\end{figure}

Here we propose a novel solution to this practical challenge by introducing the notion of \emph{Projected topological branes} (PTBs): Holographic images of higher-dimension topological crystals on lower-dimensional branes. Depending on the orientation of such branes in the parent crystal, they can either manifest an emergent lattice periodicity or aperiodic quasicrystals, as illustrated in Fig.~\ref{fig:projectiongeometry}. The latter  give birth to \emph{topoquasicrystals} within the larger territory of PTBs. In particular, the PTBs are constructed by implementing the top-bottom geometric cut-and-project procedure on the lattice models of the parent topological crystals. Remarkably, the PTBs inherit and manifest the key topological features of parent crystals: the bulk-boundary [Fig.~\ref{fig:endpointLDOS}] and the bulk-lattice defect [Figs.~\ref{fig:dislocationgeometry} and~\ref{fig:dislocationLDOS}] correspondences and topological invariant [Fig.~\ref{fig:topoinvariant}] for insulators. Topological semimetals on PTBs inherit the bulk-boundary correspondence [Fig.~\ref{fig:projectedWSM}] and topological response [Fig.~\ref{fig:Chiralanomaly}] of the parent phase.

\noindent
{\bf General framework}.~PTBs are described by the Hamiltonian for sharp quasiparticles with infinite lifetime
\allowdisplaybreaks[4]	
\begin{equation}~\label{eq:generalHamil}
H_{\rm PTB} = H_{11} - H_{12} H_{22}^{-1} H_{21},
\end{equation}
realized by integrating out the sites living outside the lower-dimensional brane.~Here $H_{11}$ and $H_{22}$ are the Hamiltonian for the sites residing within and falling outside such brane, respectively, and $H_{12}$ captures the coupling between them, with $H_{21} = H_{12}^\dagger$. The Hamiltonian for the parent crystal thus takes the block form
\begin{equation}~\label{eq:parentHamilblock}
H_{\rm parent}= \left( \begin{array}{cc}
H_{11} & H_{12} \\
H_{21} & H_{22}
\end{array}
\right).
\end{equation}
By construction, this framework is insensitive to the dimensionality and symmetry class of the parent system, the range of hopping and the slope of the brane therein, as long as the inverse of $H_{22}$ exists. The latter condition can be satisfied as possible singularities (zero modes of $H_{22}$) are always \emph{isolated}, and therefore can be regularized by taking a proper limiting procedure~\cite{silvester-2000}. See Supplementary Note 1 of the Supplemental Information (SI). Therefore, gapped phases on the parent crystal yield gapped states on the branes, while here we conjecture that the PTBs possibly inherit gapless topology from the parent gapless phase, as a specific example with isolated point nodes (Weyl semimetal) suggests [Figs.~\ref{fig:projectedWSM} and ~\ref{fig:Chiralanomaly}]. Notice that both $H_{\rm parent}$ and $H_{\rm PTB}$ are Hermitian, thereby always yielding real energy eigenvalue spectra.

From the structural point of view, when such a brane is inclined with a rational and irrational slope within the parent crystal, the projected lattice constitutes a lower-dimensional crystal with an emergent periodicity and a quasicrystal, devoid of the translational symmetry, respectively~\cite{goldman-rmp1993, jagannathan-rmp2021}. If the parent Hamiltonian ($H_{\rm parent}$) contains only nearest-neighbor hopping elements, each of the components is schematically shown in Fig.~\ref{fig:projectiongeometry}.

In the context of quasicrystals, here we bridge a long-standing gap between the description of their quantum-mechanical electronic and geometric  properties by introducing the notion of \emph{topoquasicrystal} that  realizes the dimensional descendent in the Hilbert space of the electronic wavefunctions through the projection of the parent Hamiltonian onto the quasicrystalline brane. This approach is fundamentally different than earlier works, either implementing specific topological models directly on quasicrystal networks~\cite{tran-prb2015,huang-prl2018,huang-prb2018,chen-prl2020} or allowing access to specific quasicrystals in the presence of aperiodic potentials~\cite{kraus-prl2012-1, kraus-prl2012-2, ganeshan-prl2013, kraus-prl2013}. Most importantly, as we demonstrate here, both crystal and quasicrystal inherit the electronic topology of the parent crystal, opening up a path to explore the electronic topology of higher-dimensional crystals, otherwise inaccessible in laboratory.

\noindent
{\bf Key results}.~Here we illustrate generic features of PTBs by focusing on a paradigmatic and possibly the simplest example:  parent 2D Chern insulator on a square lattice. By implementing the geometric cut-and-project procedure on the corresponding lattice regularized Hamiltonian, we showcase its footprint on an emergent 1D crystal as well as on the 1D Fibonacci quasicrystal. In particular, we find that in the entire parameter regime of the parent 2D Chern insulator, the projected 1D chain hosts topological endpoint modes [Fig.~\ref{fig:endpointLDOS}]. In addition, when the 2D Chern insulator is translationally active, featuring finite momentum band inversion, e.g., at the ${\rm M}$ point of the 2D Brillouin zone, dislocation lattice defects harbor robust topological modes around its core, as long as the defect core is confined within the projected 1D chain [Figs.~\ref{fig:dislocationgeometry} and ~\ref{fig:dislocationLDOS}]. In addition, both parent 2D Chern insulator and 1D Chern insulator on PTBs possess quantized local Chern number [Fig.~\ref{fig:topoinvariant}]. Finally, by stacking such 1D projected Chern insulators we construct Weyl semimetals in two dimensions, bearing hallmarks of a parent 3D Weyl semimetal. Specifically, we show the existence of gapless Weyl nodes, Fermi arc surface states [Fig.~\ref{fig:projectedWSM}], as well as the chiral zeroth Landau level. The latter is responsible for chiral anomaly, which we anchor by numerically computing the associated universal coefficient [Fig.~\ref{fig:Chiralanomaly}].

\noindent
{\bf Results}\\
{\bf Model}.~We consider the spinless Bernevig-Hughes-Zhang model for 2D insulators~\cite{bhz-science2006}
\begin{equation}~\label{eq:hamilchern}
H=t \sum_{j=x,y} \sin(k_j a) \tau_j
-\bigg[ m-2 t_0 \sum_{j=x,y}\left[ 1- \cos(k_j a) \right] \bigg] \tau_z.
\end{equation}
Here $\vec{k}=(k_x,k_y)$ is the spatial momentum and $a$ is the lattice spacing. The vector Pauli matrices $\boldsymbol{\tau}=(\tau_x,\tau_y,\tau_z)$ operate on sublattice index. This model harbors topological insulator for $0<m/t_0<8$. Otherwise, the system describes a trivial or normal insulator. Within the former regime, there are two distinct phases. Namely, (a) the $\Gamma$ phase for $0<m/t_0<4$, with the band inversion at the $\Gamma=(0,0)$ point of the Brillouin zone, and (b) the ${\rm M}$ phase for $4<m/t_0<8$, featuring the band inversion at the ${\rm M}=(\pi,\pi)/a$ point. These two phases are characterized by the nontrivial bulk topological invariant, the first Chern number $C=-1$ and $+1$, respectively, while $C=0$ for normal insulator. See Supplementary Note 2A and Supplementary Figure 1 of the SI.

{\bf Bulk-boundary correspondence}.~We numerically diagonalize $H$ [Eq.~(\ref{eq:hamilchern})] upon implementing it on a square lattice with open boundary conditions, which yields one-dimensional edge modes only when $0<m/t_0<8$, manifesting the bulk-boundary correspondence. Next, we construct the effective Hamiltonian ($H_{\rm PTB}$) following the prescription shown in Eq.~(\ref{eq:generalHamil}) on the 1D chain, embedded in the parent square-lattice crystal. See Supplementary Note 2B of the SI. In the entire topological regime we find near zero-energy in-gap modes, which are highly localized at the end points of such a 1D chain. By contrast, in the trivial regime the spectra of $H_{\rm PTB}$ are devoid of such endpoint modes. To anchor this claim, we repeat this procedure in a system with periodic boundary condition, for which the end-point in-gap modes therefore disappear from the spectra of $H$ and $H_{\rm PTB}$. The results are displayed in Fig.~\ref{fig:endpointLDOS}. We arrive at identical conclusions irrespective of the slope of the embedded 1D chains in the parent square lattice. These observations in turn strongly suggest the existence of insulating PTBs on 1D crystals and Fibonacci quasicrystals from the bulk-boundary correspondence, with the endpoint in-gap modes as the holographic images of the 1D edge modes of parent Chern insulators on 2D square lattice.

{\bf Dislocations}.~To further corroborate the topological nature of the 1D projected brains, we study their response to the bulk dislocation defects, known to be sensitive to the electronic topology~\cite{ran-natphys2009, juricic-prl2012, hamasaki-apl2017, nayak-sciadv2019, nag-roy-commphys2021, panigrahi-PRB2022}. In particular,  despite supporting robust gapless edge modes, the $\Gamma$ and ${\rm M}$ phases are distinguishable by the first Chern number. Furthermore, the band inversion is at different momentum (${\bf K}_{\rm inv}$) in the two phases. Therefore, the bulk dislocation lattice defects, being sensitive to ${\bf K}_{\rm inv}$, are instrumental to distinguish them.

A dislocation on the parent 2D square lattice is created by removing a line of atoms up to a site, known as its core or center, and subsequently joining the sites across the missing line of atoms: Volterra cut-and-paste procedure. As a result, any closed loop around the dislocation core features a missing translation by the Burgers vector ${\bf b}$, with ${\bf b}=a {\bf e}_x$ in Fig.~\ref{fig:dislocationgeometry}. See also Supplementary Figure 2 of the SI. Consequently, an electron in a topological insulator encircling the dislocation core picks up a hopping phase $\exp[i \Phi]$, where $\Phi={\bf K}_{\rm inv} \cdot {\bf b}$ (modulo $2\pi$)~\cite{ran-natphys2009}. Thus in the $\Gamma$ phase $\Phi=0$, while $\Phi=\pi$ in the ${\rm M}$ phase, implying that  the dislocation defect hosts a zero energy midgap state in the ${\rm M}$ phase~\cite{juricic-prl2012}. As dislocation defects are associated with the breaking of lattice translational symmetry, the ${\rm M}$ phase is referred to as translationally active~\cite{slager-natphys2013}.

Next we numerically diagonalize the projected Hamiltonian $H_{\rm PTB}$ [Eq.~(\ref{eq:generalHamil})] in the presence of a single dislocation on the parent 2D square lattice, such that the core or center of the defect resides within the embedded 1D chain [Fig.~\ref{fig:dislocationgeometry}]. See Supplementary Note 3 of the SI. Irrespective of its slope, we find zero-energy dislocation mode in the ${\rm M}$ phase of the 2D Chern insulator. The local density of states of the dislocation mode is strongly localized around the center of this defect even when projected onto the 1D chain. Nor the $\Gamma$ phase neither the trivial insulator accommodate zero-energy dislocation modes. See Fig.~\ref{fig:dislocationLDOS}.

A comment is due at this stage. Recall Fibonacci quasicrystal is devoid of translational symmetry. Still it continues to accommodate robust dislocation modes when the ${\rm M}$ phase is projected onto the quasicrystalline brane. By contrast, when the Hamiltonian in Eq.~(\ref{eq:hamilchern}) is directly implemented on an amorphous or a 2D quasicrystalline network, both of which lack the translational symmetry, the system only features a Chern insulator with $C=-1$, the shadow of the $\Gamma$ phase, and a normal insulator with $C=0$~\cite{agarwalashenoy}. However, the footprint of the ${\rm M}$ phase disappears. In this regard, 1D PTBs are genuine holographic images of the 2D Chern insulator, harnessing all the phases of the original topological crystal with dislocations as the smoking gun probe of translationally active phase therein, as long as its core resides within the brane.

{\bf Topological invariant of PTBs}.~To establish a one-to-one correspondence between the parent 2D Chern insulator on a square lattice and its realizations on 1D PTBs, finally we show that both of them possess the same topological invariant, namely the on site or local Chern number ($C_L$)~\cite{BiancoResta}. The details are discussed in Supplementary Note 2C and 2D of the SI. The local Chern number is quantized to $C_L=-1$ and $+1$ respectively for $0<m/t_0<4$ and $4<m/t_0<8$, when computed on the sites that are buried in the interior of the square lattice. On 1D PTBs, $C_L$ also takes the same quantized values, when computed on the sites that are residing at middle of two confining lines, defining the boundaries of the PTBs within the parent square lattice. These sites are shown in blue, which fall on a line shown in green in Fig.~\ref{fig:topoinvariant}(a) and~\ref{fig:topoinvariant}(c). The variation of $C_L$ along these lines is shown in Fig.~\ref{fig:topoinvariant}(b) and~\ref{fig:topoinvariant}(d), respectively when the PTB is inclined at a rational and an irrational slope, respectively. For additional results, see Supplementary Figure 1 of the SI.

{\bf Weyl branes}.~Translationally invariant stacking of topological insulators in the out of the hyperplane direction give rise to weak topological phases. Generically, they inherit the topological invariant of the underlying insulator. As such weak topological phases can be insulating or gapless, with 3D Weyl semimetal as its prominent representative~\cite{armitage-rmp2018}, which we now use to illustrate a rather simple example of a weak PTB.

Weyl semimetal can be constructed by stacking 2D Chern insulators on a cubic crystal, captured by the Hamiltonian $H_{\rm weak}=H+H_z$, where
\begin{equation}~\label{eq:weakcoupling}
H_z=t_z \cos(k_z a) \tau_z
\end{equation}
and $t_z$ is the interlayer hopping in the $z$ direction. This model accommodates both Weyl semimetals and weak Chern insulators in the $(m/t_0, t_z/t_0)$ plane~\cite{royslagerjuricic:globalPD}. For example, we find Weyl semimetals with two Weyl nodes at $\vec{k}=\vec{k}^\star_\pm$, where $\vec{k}^\star_\pm=(\pi/a,\pi/a,\pm k^\star_z)$, with $k^\star_z=\pi/(2a)$ for $m=4$ and $t=t_z=2t_0=1$. See Supplementary Figure 3 of the SI. Next we search for analogous Weyl systems in two spatial dimensions by stacking 1D projected Chern insulators in the $z$ direction, thus named projected Weyl semimetal (PWSM) hereafter.

Treating $k_z$ as a good quantum number, we construct $H_{\rm PTB} (k_z)$ from the parent Hamiltonian $H_{\rm weak}$ following the prescription from Eq.~(\ref{eq:generalHamil}), with periodic boundary conditions in the $x$ and $y$ directions. See Supplementary Note 4 of the SI. The resulting energy spectra display two Weyl nodes at $k_z=\pm k^\star_z$ for $m=4$ and $t=t_z=2t_0=1$. Thus 2D PWSM hosts Weyl nodes exactly at the same points in the 1D Brillouin zone as its parent 3D counterpart [Figs.~\ref{fig:projectedWSM}(a) and (d)].

{\bf Fermi arcs}.~The bulk-boundary correspondence of Weyl semimetal is encoded by the Fermi arc states between two Weyl nodes~\cite{armitage-rmp2018}. See Supplementary Figure 4 of the SI. Notice a PWSM is constructed by stacking 1D projected Chern insulators between the Weyl nodes ($|k_z| \leq k^\star_z$), beyond which trivial insulators occupy the 1D Brillouin zone along $k_z$. Each copy of such 1D projected Chern insulator supports endpoint modes, a locus of which constitute the Fermi arc states, when $|k_z| \leq k^\star_z$. Also notice that for each $k_z$, the localization length of the endpoint modes is inversely proportional to the bulk gap of the underlying 1D projected Chern insulator. The latter is maximum for $k_z=0$ and decreases symmetrically as we approach the Weyl nodes at $k_z=\pm k^\star_z$, where it vanishes. Consequently, the Fermi arc is maximally localized at the end points of the underlying 1D chain for $k_z=0$, while it is completely delocalized and leaks through the bulk Weyl nodes at $k_z=\pm k^\star_z$, where opposite ends of the 1D chain of the PWSM get connected. These features are shown in Fig.~\ref{fig:projectedWSM}(b) and (e). Extended Fermi arcs are also observed, when we implement open boundary condition along all three directions. See Supplementary Note 4B of the SI. We show their localization along the brane in each 2D layer in Fig.~\ref{fig:projectedWSM}(c) and (f). These outcomes are qualitatively insensitive to whether 1D chain displays crystalline or quasicrystalline order.

{\bf Chiral anomaly}.~Yet another hallmark of 3D Weyl semimetals is the chiral anomaly, stemming from the 1D chiral zeroth Landau level~\cite{nielsenninomiya}. Specifically, when a Weyl semimetal is immersed in a quantizing magnetic field (${\bf B}$) the electronic bands quench onto a set of Landau levels that are dispersive in the field direction (say $\hat{z}$), along which momentum ($k_z$) is a conserved quantity. Of particular interest is the zeroth one, the only Landau level that crosses the zero energy at the Weyl nodes. See Supplementary Figure 5 of the SI. Then in the presence of an external electric field (${\bf E}$), applied in the direction of the ${\bf B}$ field, the zeroth Landau level causes pumping of electric charge from the left to the right chiral Weyl nodes, respectively located at $\vec{k}^\star_\pm$, for example. Consequently, the number of the left and right chiral fermions is not conserved individually, leading to the celebrated \emph{chiral anomaly} in quantum crystals~\cite{nielsenninomiya, bell-jackiw, adler}. The total charge, however, remains conserved.

Next we show that the concept of chiral anomaly remains equally operative in PWSM, as it continues to host chiral zeroth Landau level crossing the zero energy at $\pm k^\star_z$. To this end we introduce a uniform magnetic field in a 3D Weyl semimetal via the Peierls substitution $\exp[\frac{2\pi i}{\Phi_0} \int^{{\bf r}_j}_{{\bf r}_i} {\bf A} \cdot d{\bf r}]$ to the hopping amplitudes between the sites at ${\bf r}_i$ and ${\bf r}_j$, where $\Phi_0=2\pi \hbar/e$ is the magnetic flux quantum, $\hbar=1$ is the Planck's constant, ${\bf A}=-B y \hat{x}$ is the vector potential, and ${\bf B}={\boldsymbol \nabla} \times {\bf A}=B \hat{z}$.
We impose periodic (open) boundary condition in the $x$ ($y$) direction and set $B=1/L_y$, where $L_y$ is the linear dimension of the system in the $y$ direction. Subsequently, we construct the effective Hamiltonian for PWSM $H_{\rm PTB}(k_z, B\hat{z})$, following Eq.~(\ref{eq:generalHamil}). See Supplementary Note 4C of the SI. Numerical diagonalization of $H_{\rm PTB}(k_z, B\hat{z})$ reveals chiral zeroth Landau level in 2D PWSM, crossing the zero energy exactly at $k_z=\pm k^\star_z$, irrespective of whether the projected 1D system features lattice periodicity or Fibonacci sequence, as respectively shown in Figs.~\ref{fig:Chiralanomaly}(a) and (c), inheriting the signatures of parent 3D Weyl semimetal. Thus 2D PWSM is expected to feature chiral anomaly.

To quantify chiral anomaly in PWSM, we note that parent time-reversal symmetry breaking 3D Weyl semimetal supports anomalous Hall conductivity, which is captured by a Chern-Simons term~\cite{zyuzinburkov, grushin, goswamitewari, vazifeh-franz, dantas-banitez-roy-surowka}. Its temporal component for a ${\bf B}$-field in the $z$ direction yields a charge density $\rho= e^2 (k^\star_z a|{\bf B}|)/(2\pi^2)$. To this end we insert a singular magnetic flux tube in the Hamiltonian for parent 3D Weyl semimetal $H_{\rm weak}(k_z)$, such that ${\bf B}=B \delta({\bf r}-{\bf r}_0)$, where ${\bf r}$ and ${\bf r}_0$ are measured in units of the lattice spacing $a$. See Supplementary Note 4D of the SI. The accumulated charge $\delta Q$ (in units of $e/\pi$) in the close vicinity of such magnetic flux tube in a 3D Weyl semimetal is predicted to be~\cite{vazifeh-franz, dantas-banitez-roy-surowka} (see Supplementary Note 5 and Supplementary Figure 6 of the SI)
\begin{equation}~\label{eq:chargeaccum}
\delta Q =Q(|{\bf B}|)-Q(0)= (k^\star_z a) \: (\Phi/\Phi_0).
\end{equation}
Next we project $H_{\rm weak}(k_z)$ to construct the corresponding $H_{\rm PTB}$, such that the core of the flux tube resides within the underlying projected 1D chain. We compute $\delta Q$ in a 2D PWSM for $\Phi/\Phi_0 \ll 1$, such that the magnetic length is sufficiently large and field theoretic predictions remain operative. The results are shown in Figs.~\ref{fig:Chiralanomaly}(b) and (d), respectively when the underlying 1D chain shows lattice periodicity and Fibonacci sequence. Within the numerical accuracy the straight lines in the $(\Phi/\Phi_0,\delta Q)$ plane display a slope equal to $\pi/2$, confirming the chiral anomaly in 2D PWSMs, with its universal coefficient quantitatively matching the field theoretic prediction.


\noindent
{\bf Discussions}. \\
Considering simple, but paradigmatic representatives of topological insulators and semimetals, namely 2D Chern insulator and 3D Weyl semimetal on parent crystals, here we demonstrate their incarnations as, respectively, 1D and 2D PTBs that feature either emergent crystalline or aperiodic quasicrystalline order. Most importantly, the proposed mechanism is sufficiently general to open up a vast unexplored territory of higher-dimensional ($d>3$) topological phases that can now be harnessed in three-dimensional world. For example, topological phases defined on five- and six-dimensional cubic lattices can be realized on 2D Penrose and 3D icosahedral topoquasicrystals, respectively. This is so because the construction of the effective Hamiltonian $H_{\rm PTB}$ is insensitive to the dimensionality of the parent topological crystal and PTB. As such, the concept of PTBs should motivate systematic symmetry-based investigations of electronic band structures in $d>3$. The proposed methodology should also be applicable to driven or Floquet crystals, where similar outcomes will lead to the construction of projected Floquet topological branes, hosting nondissipative topological quasimodes on lower-dimensional dynamic branes. Additionally, this construction can be useful to study the interaction and/or disorder effects on higher-dimensional crystals by focusing on their lower-dimensional branes. These novel avenues will be explored systematically in the future.

As far as the experimental realizations of PTBs are concerned, the designer electronic materials stand as a promising platform~\cite{manoharan-nature2012, gomesQC-NatCom2017, kempkes-natmat2019, Kempkes-natphys2019}, where our proposal can be tested by tailoring requisite hopping elements of $H_{\rm PTB}$. Robust topological modes on designer PTBs can be detected using a scanning tunneling microscope. Moreover, the projected Hamiltonian $H_{\rm PTB}$ is equally germane to classical systems. Therefore, PTBs can also be engineered on various metamaterials, among which topolectric circuits~\cite{ninguyan-prx2015,albert-prl2015,imhof-natphys2018}, phononic~\cite{susstrunk-science2015,yang-prl2015} and photonic~\cite{ozawa-rmp2019} lattices are the most promising ones. Topological modes on projected circuit branes can be detected from a diverging electrical impedance, while those in projected photonic (phononic) branes can be identified from two-point pump probe or reflection spectroscopy (mechanical susceptibility). In the future it will be worthwhile to investigate the robustness of PTBs on the range of hopping therein, given that longer-range hopping elements are generated by the projection. Such analysis will further increase the prospect of material realizations of PTBs at least in designer and metamaterials. Indeed, longer range hoppings have recently been engineered on topolectric circuits~\cite{olekhno-prb2022}, which can possibly be also implemented on photonic and phononic lattices.

\noindent
{\bf Methods}\\
To capture topological properties of PTBs featuring either emergent crystalline or quasicrystalline Fibonacci order [Fig.~\ref{fig:projectiongeometry}], we compute the corresponding effective Hamiltonian ($H_{\rm PTB}$) from the parent higher-dimensional crystal, described by the Hamiltonian $H_{\rm parent}$. This procedure is summarized in Eqs.~(\ref{eq:generalHamil}) and (\ref{eq:parentHamilblock}). For the sake of concreteness, here we focus on 2D square lattice quantum anomalous Hall insulator and capture its footprints on 1D PTBs. Then from the energy spectra and the spatial distribution of the near zero energy modes of $H_{\rm PTB}$ we demonstrate the bulk-boundary [Fig.~\ref{fig:endpointLDOS}] and bulk-dislocation [Fig.~\ref{fig:dislocationLDOS}] correspondences. For the latter, we ensure that the dislocation core resides within the PTB [Fig.~\ref{fig:dislocationgeometry}]. The local Chern number for PTBs is computed from the eigenfunctions of all the filled states at negative eigenenergies of $H_{\rm PTB}$ [Fig.~\ref{fig:topoinvariant}]. While constructing the effective Hamiltonian for projected Weyl branes in two dimensions from the parent 3D Weyl semimetals, realized by stacking layers of 2D Chern insulators in the $z$ direction in a translationally invariant fashion, we treat $k_z$ as good quantum number and therefore the projection is performed only on the $xy$ plane. This procedure is employed to compute the effective band structure and Fermi arcs [Fig.~\ref{fig:projectedWSM}], as well as Landau level spectra and chiral anomaly [Fig.~\ref{fig:Chiralanomaly}] of 2D projected Weyl branes. To compute the universal coefficient associated with the chiral anomaly, we ensure that the core of the magnetic flux tube threads the PTB. To capture the real space structure of the Fermi arcs [Fig.~\ref{fig:projectedWSM}], we also consider a parent 3D system with finite linear dimension in the $z$ direction. Still then we perform the projection only on the $xy$ plane to construct 2D Weyl brane. Additional technical details for each scenario are discussed in the Supplemental Information. \\

\noindent
{\bf Data availability}  \\
The data and software code for generating the figures presented in the main text and supplementary materials are available at https://doi.org/10.5281/zenodo.6851134.

\noindent
{\bf Acknowledgment} \\
A.P. acknowledges support from the KVPY program, and thanks Arghya Ranjan Das for technical support. V.J. acknowledges the support from the Swedish Research Council (VR 2019-04735). B.R. was supported by the Startup Grant from Lehigh University. Nordita is supported in part by Nordforsk.

\noindent
{\bf Author contributions} \\
A.P. performed all the numerical calculations. V.J. and B. R. conceived and structured the project, performed analytical calculations and wrote the manuscript. B. R. supervised the project.

\noindent
{\bf Competing interests} \\
The authors declare no conflicts of interest.


\end{document}